\documentclass{article}
 
\usepackage{amsfonts}
\usepackage{amsthm}
\usepackage{amsmath}
\usepackage{fullpage}

\newtheorem{lemma}{Lemma}[section]
\newtheorem{proposition}{Proposition}[section]
\newtheorem{theorem}{Theorem}[section]

\theoremstyle{remark}
\newtheorem{remark}{Remark}[section]
\theoremstyle{definition}

\title{Spectral Properties and Linear Stability of Self--Similar Wave Maps}
\author{Roland Donninger\thanks{\tt roland.donninger@univie.ac.at}{ }
and Peter C. Aichelburg\thanks{\tt aichelp8@univie.ac.at}\\
\small{Faculty of Physics, Gravitational Physics} \\ 
\small{University of Vienna} \\ 
\small{Boltzmanngasse 5, A-1090 Wien}}

\begin{document}

\maketitle

\abstract{
We study co--rotational wave maps from $(3+1)$--Minkowski space to 
the three--sphere $S^3$.
It is known that there exists a countable family $\{f_n\}$ of self--similar 
solutions.
We investigate their stability under linear perturbations by 
operator theoretic methods.
To this end we study the spectra of the perturbation operators, prove
well--posedness of the corresponding linear Cauchy problem and deduce a growth
estimate for solutions.
Finally, we study perturbations of the limiting solution which is obtained from
$f_n$ by letting $n \to \infty$.}

\section{Introduction and Definition of the Model}
Many nonlinear evolution equations show a behaviour known as "blow up": 
Although the initial
data are regular, the solution becomes singular in finite time.
Important questions in this respect are: Under what circumstances does the blow
up occur? What is the mechanism that leads to singularity formation? What is the
blow up rate and the shape of the profile before the singularity forms?
Unfortunately, many interesting examples from physics 
(such as Einstein's equations)
are much too complicated to be accessible to presently available analytical
techniques.
Thus, one has to rely on numerical studies and simplified "toy models" which share
some features with the original equation but are easier to handle.

The wave maps system provides an example of such a toy model which has been 
studied in recent times.
Consider ($n+1$)--dimensional Minkowski space $(\mathbb{R}^{n+1}, \eta)$ with
metric $(\eta_{\mu \nu})=\mathrm{diag} (-1,1, \dots, 1)$, an 
$m$--dimensional Riemannian manifold $(M,g)$ with metric
$g$ and a mapping $\Phi: \mathbb{R}^{n+1} \to M$.
Choosing local coordinates on $M$ we denote the components of $\Phi$ and the
metric $g$ by $\Phi^A$ and $g_{AB}$, respectively, 
where capital latin indices run from $1$ to $m$.
The action functional $S$ for wave maps is defined by
$$ S(\Phi)=\int_{\mathbb{R}^{n+1}} \eta^{\mu \nu} (\partial_\mu \Phi^A)
(\partial_\nu \Phi^B) g_{AB} \circ \Phi $$
where integration is understood with respect to the ordinary Lebesgue measure on
$\mathbb{R}^{n+1}$ and Einstein's summation convention is in force (greek
indices run from $0$ to $n$).
As usual we write $\partial_\mu:=\frac{\partial}{\partial x^\mu}$.
The Euler--Lagrange equations associated to this action are given by
\begin{equation}
\label{int_eq_wmsys}
\Box \Phi^A + \eta^{\mu \nu} \Gamma^A{}_{BC}(\Phi)(\partial_\mu \Phi^B)
(\partial_\nu \Phi^C)=0
\end{equation}
where $\Gamma^A{}_{BC}:=\frac{1}{2}g^{AD}(\partial_B g_{CD}+\partial_C
g_{BD}-\partial_D g_{BC})$ are the Christoffel symbols associated to the metric
$g$ on $M$ and $\Box:=\eta^{\mu \nu}\partial_\mu \partial_\nu$ denotes the wave
operator.
Eq. (\ref{int_eq_wmsys}) is known as the \emph{wave maps equation} and it is a
system of semilinear wave equations. 

We are interested in the Cauchy problem for eq. (\ref{int_eq_wmsys}), i.e. we
prescribe initial data $\Phi^A|_{t=0}=\Phi_0^A$ and $\partial_t \Phi^A
|_{t=0}=\Phi_1^A$ where $\partial_t:=\partial_0$ and study the future development.
There exists a well--established local well--posedness theory for the wave maps
system which states that for any initial data $(\Phi_0^A, \Phi_1^A) \in
H^s \times H^{s-1}$ ($s>\frac{n}{2}$) there exists a
$T>0$ and functions $\Phi^A \in C([0,T],H^s)$ with $\partial_t
\Phi^A \in C([0,T],H^{s-1})$ that solve eq. (\ref{int_eq_wmsys})
in the sense of distributions and satisfy 
$(\Phi^A, \partial_t \Phi^A)|_{t=0}=(\Phi_0^A, \Phi_1^A)$ where $H^s:=
H^s(\mathbb{R}^n)$ denotes the usual fractional Sobolev space.
Moreover, the time $T$ can be chosen to depend continuously on the initial data
and the mapping $(\Phi_0^A, \Phi_1^A) \mapsto (\Phi^A, \partial_t \Phi^A):
H^s \times H^{s-1} \to C([0,T],H^s) \times C([0,T],H^{s-1})$ is continuous (see
\cite{Klainerman2002} and references therein for details). 
There are also many recent results addressing global issues, see e.g.
\cite{Tataru2004} for a review.
Of most interest for our purposes is Tao's work \cite{Tao2001} which deals with
spherical targets, i.e. $M$ is an $(m-1)$--sphere for $m,n \geq 2$.
Under this assumption a global smooth solution of the wave maps system eq.
(\ref{int_eq_wmsys}) exists whenever the initial data are smooth and small in
the homogeneous Sobolev space $\dot{H}^{n/2} \times \dot{H}^{n/2-1}$. 

The present work deals with co--rotational wave maps from ($3+1$)--Minkowski 
space to the
three--sphere $S^3$.
Choosing the usual spherical coordinates $(t,r,\theta,\varphi)$ on Minkowski space and
standard coordinates $(\psi, \Theta, \phi)$ on $S^3$ with metric
$g=d\psi^2+\sin^2 \psi (d\Theta^2+\sin^2 \Theta d\phi^2)$, we restrict 
ourselves to mappings of the form $(t,r,\theta,\varphi) \mapsto (\psi(t,r),\theta, \varphi)$ which are called
\emph{co--rotational}.
Under these assumptions the system eq. (\ref{int_eq_wmsys}) reduces to the 
single semilinear wave equation
\begin{equation}
\label{int_eq_wm}
\psi_{tt}-\psi_{rr}-\frac{2}{r}\psi_r + \frac{\sin (2\psi)}{r^2}=0.
\end{equation}
We remark that Tao's result applies to this problem but there are earlier papers
by Sideris \cite{Sideris1989} and Shatah, Tahvildar--Zadeh 
\cite{Shatah1994}
which address global questions for this particular model.

\section{Self--Similar Solutions and Singularity Formation}
Note that eq. (\ref{int_eq_wm}) is invariant under dilations, i.e. if $\psi$ is
a solution, so is $\psi_\lambda$ defined by $\psi_\lambda(t,r):=\psi(t/\lambda,
r/\lambda)$ for a $\lambda>0$.
There exists a conserved energy associated to eq. (\ref{int_eq_wm}) given by
$$ E_\psi(t)=\int_0^\infty \left (r^2 \psi_t^2(t,r)+r^2 \psi_r^2(t,r)+2 \sin^2
(\psi(t,r)) \right )dr $$
that scales as $E_{\psi_\lambda}=\lambda E_\psi$ which means that
eq. (\ref{int_eq_wm}) is \emph{energy supercritical}.
It is widely believed that for energy supercritical equations
blow up is possible since
it is energetically favourable for solutions to shrink which might result in the
formation of a singularity.
In order to obtain an explicit example of a blow up solution it is reasonable to
look for self--similar solutions $\psi(t,r)=f(\frac{r}{T-t})$ where $T>0$ is a
parameter (the \emph{blow up time}).
Plugging this ansatz into eq. (\ref{int_eq_wm}) yields
\begin{equation}
\label{ss_eq_css}
f''+\frac{2}{r}f'-\frac{\sin(2f)}{\rho^2(1-\rho^2)}=0
\end{equation}
where $':=\frac{d}{d\rho}$ and $\rho:=\frac{r}{T-t}$.
We require the regularity conditions $f(0)=0$ and $f(1)=\frac{\pi}{2}$.
Shatah \cite{Shatah1988} has shown existence of a smooth solution $f_0$ of eq.
(\ref{ss_eq_css})
which has later been found in closed form \cite{Turok1990}
and it is
given by $f_0(\rho)=2 \arctan(\rho)$.
Bizo\'n \cite{Bizon2000a} has extended this result by proving existence 
of a countable
family $\{f_n: n=0,1,\dots\}$ of analytic solutions of eq. (\ref{ss_eq_css}) 
satisfying $f_n(0)=0$ and
$f_n(1)=\frac{\pi}{2}$ where $n$ counts the number of intersections of $f_n$
with the line $\frac{\pi}{2}$ on the interval $[0,1)$.
Furthermore, the solutions $f_n$ converge to the constant $f_\infty \equiv
\frac{\pi}{2}$ for $n \to \infty$ pointwise on $(0,1]$. 
These self--similar wave maps provide explicit examples of solutions
with smooth finite energy initial data (use appropriate cut--off functions and
finite speed of propagation) that become singular when $t \to T-$.
 
An obviously important question is whether blow up occurs for generic initial
data or the self--similar examples are "exceptional cases".
This issue has been addressed by Bizo\'n et. al. \cite{Bizon2000} using
numerical techniques.
Their results strongly suggest that the self--similar solution $f_0$ provides a
universal blow up profile.
More precise, they conjecture that there exists a large set of initial data 
that lead to blow up
and the singularity formation takes place via $f_0$, i.e. the time evolution
approaches $f_0$ near the center $r=0$ as $t \to T-$.   

We remark that this self--similar blow up behaviour is very different from
singularity formation in the 
energy critical case of co--rotational wave maps from (2+1)--Minkowski
space to the two--sphere.
For this model it has been known for quite some time (see \cite{struwe}) that
the blow up rate cannot be self--similar. 
However, rigorous existence of blow up solutions has only recently been shown by
Rodnianski and Sterbenz \cite{rodnianski} as well as Krieger, Schlag,
Tataru \cite{krieger}.

\section{The Linear Stability Problem}
As mentioned, the self--similar solution $f_0$ seems to play an important role in
dynamical time evolution.
Thus, it has to be stable in a certain sense since otherwise it could not be 
found by numerical time evolutions starting from generic initial data.
In order to study linear stability of the wave map $f_n$ we introduce adapted
coordinates $(\sigma, \rho)$ given by $\sigma=-\log \sqrt{(T-t)^2-r^2}$ and
$\rho=\frac{r}{T-t}$ (cf. \cite{Bizon2000a}).
The introduction of similarity coordinates in the study of self--similar blow up
is very common (see e.g. \cite{merle}).
We refer to $(\sigma,\rho)$ as \emph{hyperbolic coordinates} since the lines
$\sigma=const$ are hyperbolae in a spacetime diagram.
Note that the coordinate system $(\sigma, \rho)$ is only defined for $r<T-t$ and
thus, it covers the interior of the
backward lightcone of the blow up point $(t,r)=(T,0)$.  
Furthermore, $(\sigma, \rho)$ is an orthogonal coordinate system, i.e. the
vectorfields $\partial_\sigma$ and $\partial_\rho$ are orthogonal with respect
to the Minkowski metric.
Moreover, the blow up time $t=T$ is shifted to infinity, i.e. $t \to T-$ 
corresponds to
$\sigma \to \infty$.
The valid range is $\sigma \in \mathbb{R}$ and $\rho \in [0,1)$.
Eq. (\ref{int_eq_wm}) transforms into
\begin{equation}
\label{lin_eq_wmhyp}
\psi_{\sigma \sigma}-2 \psi_\sigma - (1-\rho^2)^2 \psi_{\rho
\rho}-\frac{2(1-\rho^2)^2}{\rho}\psi_\rho+\frac{(1-\rho^2)\sin(2
\psi)}{\rho^2}=0.
\end{equation}
Plugging the ansatz $\psi(\sigma,\rho)=f_n(\rho)+\phi(\sigma,\rho)$ into eq.
(\ref{lin_eq_wmhyp}) and
linearizing around $f_n$ we arrive at the evolution equation
\begin{equation}
\label{lin_eq_phi} \phi_{\sigma \sigma}-2 \phi_\sigma - (1-\rho^2)^2 \phi_{\rho
\rho}-\frac{2(1-\rho^2)^2}{\rho}\phi_\rho+\frac{2(1-\rho^2)\cos(2
f_n)}{\rho^2}\phi=0 
\end{equation}
for the perturbation $\phi$ which governs the linearized flow around the wave
map $f_n$.
The wave map $f_n$ is considered to be linearly stable if there do not exist
solutions of eq. (\ref{lin_eq_phi}) that grow (with respect to a suitable
norm) as $\sigma$ increases.
We will make this more precise below.
However, a direct computation shows that $\phi^G_n$ defined by $\phi^G_n(\sigma,
\rho):=e^\sigma \rho \sqrt{1-\rho^2}f_n'(\rho)$ solves eq. (\ref{lin_eq_phi}).
This instability is related to the time translation symmetry of the original
problem.
The reason for this is the fact that $f_n$ is not a single solution but a 
one--parameter family
of solutions depending on the blow up time $T$.
We will refer to this instability as the \emph{gauge instability}.
Thus,
without further considerations the best we can expect is to prove a result that
excludes the existence of solutions of
eq. (\ref{lin_eq_phi}) with $f_n=f_0$ that grow \emph{faster} than this gauge 
instability. 
In what follows we will address this issue by using an operator theoretic
approach. 

\section{Operator Formulation}
In order to simplify the problem further we define a new unknown
$\tilde{\phi}$ by $\tilde{\phi}(\sigma, \rho):=e^{-\sigma}\phi(\sigma, \rho)$ 
to remove the first--order
term $-2 \phi_\sigma$.
Eq. (\ref{lin_eq_phi}) transforms into
\begin{equation}
\label{op_eq_tildephi}
\tilde{\phi}_{\sigma \sigma} - (1-\rho^2)^2 \tilde{\phi}_{\rho
\rho}-\frac{2(1-\rho^2)^2}{\rho}\tilde{\phi}_\rho+\frac{2(1-\rho^2)\cos(2
f_n)-\rho^2}{\rho^2}\tilde{\phi}=0.
\end{equation}

\subsection{The Operator $A_n$}
We define the formal Sturm--Liouville differential expression $a_n$ on the
interval $(0,1)$ by
$$ a_n u:=\frac{1}{w}(-(pu')'+q_nu) $$
where $w(\rho):=\frac{\rho^2}{(1-\rho^2)^2}$, $p(\rho):=\rho^2$ and
$q_n(\rho):=\frac{2(1-\rho^2)\cos(2 f_n(\rho))-\rho^2}{(1-\rho^2)^2}$.
Thus, $a_n$ represents the spatial differential operator in eq.
(\ref{op_eq_tildephi}).
We define the weighted Hilbert space $H:=L^2_w(0,1)$ where $L^2_w(0,1)$ 
consists of
(equivalence classes of) functions which are square integrable with respect to
the weight $w$, i.e. the inner product $(\cdot|\cdot)_H$ on $H$ is given by
$$ (u|v)_H=\int_0^1 u(\rho) \overline{v(\rho)} w(\rho) d\rho $$
and as usual we write $\|\cdot\|_H:=\sqrt{(\cdot | \cdot)_H}$.
We set $\mathcal{D}(A_n):=\{u \in H: u,pu' \in AC_\mathrm{loc}(0,1), a_n u \in H\}$ and
$A_n u:=a_n u$ for $u \in \mathcal{D}(A_n)$.

\begin{lemma}
The operator $A_n: \mathcal{D}(A_n) \subset H \to H$ is self--adjoint.
\end{lemma}

\begin{proof}
Consider solutions of $a_n u=0$.
Since the coefficients of this linear ordinary differential equation are
meromorphic and $\rho=0$ as well as $\rho=1$ are regular singular points, we can
apply Frobenius' method to obtain series representations for solutions around
$\rho=0$ and $\rho=1$.
Around $\rho=0$ the \emph{Frobenius indices} are $-2$ and $1$, i.e. only one
nontrivial solution belongs to $H$.
According to the Weyl alternative we conclude that $\rho=0$ is in the
limit--point case (cf. e.g. \cite{Zettl2005}).
Similarly, around $\rho=1$ the indices are both $\frac{1}{2}$ and thus, neither
of the two linearly independent solutions belongs to $H$ which shows that 
$\rho=1$
is in the limit--point case as well.
It follows that the maximal operator $A_n$ associated to $a_n$ is self--adjoint
(cf. \cite{Zettl2005}).
\end{proof}

\subsection{Factorization of $A_0$}
Very similar to the energy critical case \cite{rodnianski} there exists a
factorization of the linearized operator $A_0$ that immediately implies its 
nonnegativity.
To see this, note first that the potential term  
in eq. (\ref{lin_eq_phi}) for the ground state $f_0$ can be rewritten
according to
$$ \cos(2f_0(\rho))=\cos(4\arctan \rho)=\frac{1-6\rho^2+\rho^4}{(1+\rho^2)^2}.
$$
Now we define two first--order operators $B$ and $\hat{B}$ on $H$ by
$\mathcal{D}(B):=\mathcal{D}(\hat{B}):=C^\infty_c(0,1)$ as well as
$$Bu(\rho):=(1-\rho^2)u'(\rho)-\frac{1-3\rho^2}{\rho(1+\rho^2)}u(\rho)$$ and
$$\hat{B}u(\rho):=-(1-\rho^2)u'(\rho)-\frac{3-\rho^2}{\rho(1+\rho^2)}u(\rho).$$
A simple integration by parts shows that $(Bu|v)_H=(u|\hat{B}v)_H$ for all $u,v
\in C^\infty_c(0,1)$ and thus, $B$ and $\hat{B}$ are adjoint to each other.
Furthermore, by a direct computation one verifies that $A_0 u=\hat{B}Bu$ for all
$u \in C^\infty_c(0,1)$ which is the desired factorization.
We immediately obtain $(A_0u|u)_H=(\hat{B}Bu|u)_H=(Bu|Bu)_H\geq 0$ for all $u
\in C^\infty_c(0,1)$.
Since $A_0$ is limit--point at both endpoints it follows that $C^\infty_c(0,1)$
is a core for $A_0$ and the estimate $(A_0u|u)\geq 0$ extends to all $u \in
\mathcal{D}(A_0)$ by a density argument.
Thus, $A_0$ is nonnegative and we have $\sigma(A_0) \subset [0,\infty)$.

\subsection{The Spectrum of $A_n$}
\begin{proposition}
The point spectrum $\sigma_p(A_n)$ of $A_n$ consists of exactly $n$ negative
eigenvalues, i.e. $\sigma_p(A_n)=\{\lambda_1, \dots, \lambda_n:
0>\lambda_1 > \dots >\lambda_n\}$.
Furthermore, the essential spectrum $\sigma_e(A_n)$ is given by 
$\sigma_e(A_n)=[0,\infty)$.
\end{proposition}

\begin{proof}
Consider the equation $(\lambda-a_n)u=0$.
The Frobenius indices around $\rho=1$ are $\frac{1 \pm \sqrt{-\lambda}}{2}$.
Thus, for $\lambda \geq 0$ neither of the two linearly independent solutions
belongs to $H$ which shows $\sigma_p(A_n) \subset (-\infty, 0)$.
Furthermore, we conclude $\sigma_e(A_n)=[0,\infty)$ by \cite{weidmann}, 
p. 165, Theorem 11.5c.

As already discussed, the function $\theta_n$ defined by $\theta_n(\rho):=\rho
\sqrt{1-\rho^2}f'_n(\rho)$ is a solution of $a_n u=0$.
Bizo\'n \cite{Bizon2000a} has shown that $f_n(\rho) \in (0,\pi)$ for $\rho \in
(0,1]$ and thus, if $f_n>\frac{\pi}{2}$, it can only have one maximum
since 
$$ f''_n(\rho)=\frac{\sin(2f_n(\rho))}{\rho^2(1-\rho^2)}<0 $$
for $f'_n(\rho)=0$ (cf. eq. (\ref{ss_eq_css})).
Similarly, if $f_n<\frac{\pi}{2}$, it can only have one minimum.
Hence, the number of extrema of $f_n$ equals the number of intersections of
$f_n$ with the line $\frac{\pi}{2}$ on $(0,1)$ and therefore, $\theta_n$ has
exactly $n$ zeros on $(0,1)$.

Applying the transformation $v(x):=\sinh (x)u(\tanh x)$, the equation
$(\lambda-a_n)u=0$ reduces to the eigenvalue problem
\begin{equation}
\label{eq_schrodinger} 
\lambda v(x) + v''(x) -\frac{2}{x^2}v(x)-Q_n(x)v(x)=0 
\end{equation}
where
$$ Q_n(x):=\frac{2\cos(2f_n(\tanh x))}{\sinh^2 x}-\frac{2}{x^2}. $$
Consequently, the function $\varphi_n(x):=\sinh(x)\theta_n(\tanh x)$ is a solution of
eq. (\ref{eq_schrodinger}) with $\lambda=0$ which has exactly $n$ zeros on 
$x \in (0,\infty)$.
Observe that the mapping $u \mapsto v$ is an isometry from $L^2_w(0,1)$ to
$L^2(0,\infty)$ and thus, the study of $A_n$ is equivalent to
the spectral problem for
the $\ell=1$ radial Schr\"odinger operator
on $L^2(0,\infty)$ with potential $Q_n$.
Note that $Q_n \in L^\infty(0,\infty)$ and
from the Frobenius analysis of $a_n$ it follows that 
the asymptotic behaviour of nontrivial solutions of eq. 
(\ref{eq_schrodinger}) is the same as if $Q_n \in C^\infty_c(0,\infty)$.
Thus, the results of \cite{reedsimon}, XIII.3.B apply and we conclude that $A_n$
has exactly $n$ eigenvalues below the essential spectrum.
\end{proof}

\begin{remark}
This observation has already been made by Bizo\'n \cite{Bizon2000}.
Note further that $\theta_n$ is not an eigenfunction since it does not belong to
$H$!
\end{remark}

\begin{remark}
Negative eigenvalues of $A_n$ lead to unstable mode solutions of eq.
(\ref{lin_eq_phi}), i.e. $\phi(\sigma, \rho):=e^{(1+\sqrt{-\lambda})
\sigma} v(\rho)$ is a
solution of eq. (\ref{lin_eq_phi}) if $\lambda<0$ belongs to $\sigma_p(A_n)$
and $v$ is the
associated eigenfunction.
Thus, the $n$--th self--similar solution has exactly $n$ unstable modes.
\end{remark}

\section{Well--Posedness of the Linearized Problem and Growth of Solutions}
We consider the linear Cauchy problem
\begin{equation}
\label{wp_eq_Cauchy}
\left \{ \begin{array}{l}
\ddot{u}(\sigma)+A_0 u(\sigma)=0 \mbox{ for } \sigma > 0\\
u(0)=u_0, \dot{u}(0)=u_1
\end{array} \right .
\end{equation}
for a function $u: [0,\infty) \to H$ and 
$u_0,u_1 \in \mathcal{D}(A_0)$ where $\dot{}:=\frac{d}{d\sigma}$
and
differentiation with respect to $\sigma$ is understood in the strong sense.
Eq. (\ref{wp_eq_Cauchy}) is an operator formulation of eq. (\ref{op_eq_tildephi})
with $f_n=f_0$ and thus, it describes the (rescaled) linear flow around the 
self--similar wave
map $f_0$.

According to the spectral theorem for self--adjoint operators (see e.g.
\cite{Yosida1980}) there exists a spectral family $\{E(\lambda): \lambda \in
\mathbb{R}\}$ of orthogonal projections associated to $A_0$ such that for $u \in
\mathcal{D}(A_0)$ and $v \in H$ we have
$$ (A_0 u|v)_H=\int \mathrm{id}_\mathbb{R} d\mu_{u,v} $$
where $\mu_{u,v}:=\frac{1}{4}(\mu_{u+v}-\mu_{u-v}+i\mu_{u-iv}-i\mu_{u+iv})$
and $\mu_u$ is the unique measure defined on the Borel--$\sigma$--algebra on
$\mathbb{R}$ that satisfies
$\mu_u((\lambda_1,\lambda_2])=\|(E(\lambda_2)-E(\lambda_1))u\|_H^2$ for $u \in H$.
Moreover, with the help of the spectral family $\{E(\lambda)\}$ it is possible
to define functions of self--adjoint operators.
Let $f: \mathbb{R} \to \mathbb{C}$ be measureable and set 
$\mathcal{D}(f(A_0)):=\{u \in H: \int |f|^2
d\mu_u < \infty\}$, $(f(A_0)u|v)_H:=\int fd\mu_{u,v}$ for any $u \in \mathcal{D}(f(A_0))$
and $v \in H$.
Note that in fact only the values of $f$ on the spectrum of $A_0$ contribute.
This yields a well--defined linear 
operator $f(A_0): \mathcal{D}(f(A_0)) \subset H \to
H$ which behaves nicely (see \cite{Yosida1980} for details and a 
functional calculus).

\begin{proposition}
Let $u_0, u_1 \in \mathcal{D}(A_0)$. Then, the Cauchy problem eq.
(\ref{wp_eq_Cauchy}) has a unique solution $u: [0,\infty) \to \mathcal{D}(A_0)
\subset H$ and the real--valued function $\sigma \mapsto \|A_0^{1/2}
u(\sigma)\|_H^2+\|\dot{u}(\sigma)\|_H^2$ is constant for all $\sigma \geq 0$.
\end{proposition}

\begin{proof}
We set 
$$ g_\sigma(\lambda):=\left \{ \begin{array}{l} \sin (\sigma
\lambda)\lambda^{-1} \mbox{ for }\lambda>0 \\
\sigma \mbox{ for } \lambda \leq 0 \end{array} \right . $$
and define $u(\sigma):=\cos(\sigma A_0^{1/2})u_0+g_\sigma(A_0^{1/2})u_1$ 
via the functional calculus.
Note that $A_0^{1/2}$ is well--defined since $\sigma(A_0)=[0,\infty)$.
Applying Lebesgue's theorem on dominated convergence one immediately concludes
that $\ddot{u}(\sigma)+A_0u(\sigma)=0$ and clearly, $u$ satisfies the initial
conditions.
Furthermore, it follows easily that $u(\sigma)$ and $\dot{u}(\sigma)$ stay in 
$\mathcal{D}(A_0)$ for all $\sigma > 0$.
A direct computation (functional calculus) yields
$\|A_0^{1/2}u(\sigma)\|_H^2+\|\dot{u}(\sigma)\|_H^2=\|A_0^{1/2}u_0\|_H^2
+\|u_1\|_H^2$ which implies uniqueness as well.
\end{proof}

Note that $(u_0,u_1) \mapsto \|A_0^{1/2}u_0\|_H^2+\|u_1\|_H^2$ defines a norm on
$\mathcal{D}(A_0^{1/2}) \times H$ since $0 \notin \sigma_p(A_0^{1/2})$ and hence, 
there do not
exist growing solutions of eq. (\ref{wp_eq_Cauchy}) which is the desired result.
Thus, in view of the transformation $\phi \mapsto \tilde{\phi}$ we have shown
that there do not exist linear perturbations of $f_0$ that
grow faster than the gauge instability.

\section{Discussion}
We give a heuristic explanation why one cannot go any further in this
formulation.
This is most easily seen within a first--order formulation.
Let $u: [0,\infty) \to H$  be the unique solution of eq. (\ref{wp_eq_Cauchy}) 
with initial data $u_0, u_1 \in \mathcal{D}(A_0)$ and
set $\phi(\sigma, \cdot):=e^\sigma u(\sigma)$.
Then, $\phi$ satisfies the first--order equation
$$ \frac{d}{d\sigma}\mathbf{u}(\sigma)=L \mathbf{u}(\sigma) $$
where $\mathbf{u}(\sigma):=(\phi(\sigma, \cdot), \phi_\sigma(\sigma, \cdot))$
and
$$ L:=\left ( \begin{array}{cc} 1 & 1 \\ -A_0 & 1 \end{array} \right ). $$
$\phi$ (resp. $\mathbf{u}$) describes the linear flow around the self--similar wave map $f_0$.
Consider the operator $L$ on the Banach space $X:=\mathcal{D}(A_0^{1/2}) \times H$
with norm $\|(u_1, u_2)\|_X^2:=\|u_1\|_{A_0^{1/2}}^2+\|u_2\|_H^2$ and
$\mathcal{D}(L):=\mathcal{D}(A_0) \times \mathcal{D}(A_0^{1/2})$ where
$\|\cdot\|_{A_0^{1/2}}$ denotes the graph norm of $A_0^{1/2}$.
Then, it is easily seen that the spectrum of $L$ can be calculated from the
spectrum of $A_0$ and it is given by
$\sigma(L)=\{z \in \mathbb{C}: -(z-1)^2 \in \sigma(A_0)\}$ which shows that
$\sigma(L)=\{z \in \mathbb{C}: \mathrm{Re}(z)=1\}$.
We have already discussed the gauge instability which corresponds to the point
$1 \in \sigma(L)$.
Thus, the origin of the point $1 \in \sigma(L)$ is well--understood and in
fact, we 
are only
interested in stability of $f_0$ "modulo this gauge instability".
If $1 \in \sigma(L)$ would be an isolated point one could define a projection
operator to remove this instability.
However, the rest of the spectrum makes this impossible and in what follows
we give an intuitive 
explanation where it emanates from.
The reason for this is a serious defect of the coordinate system $(\sigma,
\rho)$.
As already mentioned, it covers only the interior of the backward lightcone of
the point $(T,0)$.
Thus, an outgoing wave packet can never leave the backward lightcone but 
cumulates
near $\rho=1$ as $\sigma$ increases.
This corresponds to an apparent instability 
which shows up in the spectrum of $L$.
Unfortunately, one cannot go around this difficulty unless one changes 
coordinates
which destroys the self--adjoint character of the problem. 
Thus, the result that there do not exist linear perturbations of $f_0$ that grow
faster than the gauge instability is the best one can have in this formulation
but still, it strongly supports the conjecture of stability of the wave map $f_0$.

\section{The Operators $A_n$ for $n \to \infty$}

The self--similar wave maps $f_n$ can be constructed numerically using a
shooting and matching technique \cite{Aminneborg1995}.
By the same method one may calculate the point spectrum of $A_n$ (cf.
\cite{Bizon2000}).
Results for $A_1, \dots, A_4$ are given in Table \ref{nlarge_tbl_ev}
where $\mu_j:=\sqrt{-\lambda_j}$.

\subsection{The Operator $A_\infty$}
We emphasize the remarkable fact that the horizontal rows in Table
\ref{nlarge_tbl_ev} seem to converge.
The question is if one can construct an operator $A_\infty$ such that the point
spectrum of $A_\infty$ is the limit of the point spectra of $A_n$ for $n \to
\infty$.
Since $f_n \to f_\infty \equiv \frac{\pi}{2}$ pointwise on $(0,1]$, the obvious
strategy is to investigate linear perturbations around $f_\infty$ described by
eq. (\ref{op_eq_tildephi}) with $f_n=f_\infty$.
Thus, we consider the Sturm--Liouville expression
$$ a_\infty u:=\frac{1}{w}(-(pu')'+q_\infty u) $$
where $q_\infty(\rho):=\frac{-2+\rho^2}{(1-\rho^2)^2}$.  
Again, we use Frobenius' method to classify the endpoints.
The behaviour of solutions of $a_\infty u=0$ around $\rho=1$ is the same as
before for $a_n$ and thus, $\rho=1$ is in the limit--point case again.
However, around $\rho=0$ the Frobenius indices are $\frac{-1 \pm i\sqrt{7}}{2}$
and thus, all solutions of $a_\infty u=0$ belong to $H$ near $\rho=0$.
According to the Weyl alternative the point $\rho=0$ is in the limit--circle
case and hence, we have to specify a boundary condition of the form
$[u,\chi]_p(0)=0$ for a fixed function $\chi$ 
in order to
obtain a self--adjoint operator (cf. \cite{Zettl2005}).
The question is how to choose $\chi$.
Bizo\'n \cite{Bizon2002} has studied an analogous problem for a 
Yang--Mills system.
Following his proposal we choose $\chi$ in such a way that $a_\infty u=0$ has a
nontrivial solution just like the gauge instability for $a_n$.
Thus, we denote by $\tilde{\chi}$ a real--valued nontrivial function satisfying
$a_\infty \tilde{\chi}=0$ with asymptotic behaviour $\tilde{\chi}(\rho) \sim
(1-\rho)^{1/2}$ for $\rho \to 1$.
Such a function exists by Frobenius' method and it is unique up to constant
multiples since the second linearly independent solution around $\rho=1$
contains a $\log$ term.    
Using an appropriate smooth cut--off function we construct a $\chi$ that 
coincides
with $\tilde{\chi}$ on $(0,\frac{1}{2})$ and satisfies
$\chi, a_\infty \chi \in H$ as well
as $\chi, p\chi' \in AC_\mathrm{loc}(0,1)$.
Then, we define 
$$\mathcal{D}(A_\infty):=\{u \in H: u,pu' \in
AC_\mathrm{loc}(0,1), a_\infty u \in H, [u,\chi]_p(0)=0\}$$
and $A_\infty u:=a_\infty u$ for $u
\in \mathcal{D}(A_\infty)$.
Invoking Sturm--Liouville theory \cite{Zettl2005} we conclude that $A_\infty:
\mathcal{D}(A_\infty) \subset H \to H$ is self--adjoint.

\subsection{The Point Spectrum of $A_\infty$}

Since the asymptotic behaviour of solutions of $(\lambda-a_\infty)u=0$ around
$\rho=1$ is the same as for $(\lambda-a_n)u=0$, we conclude that
$\sigma_p(A_\infty) \subset (-\infty, 0)$.
Thus, we restrict ourselves to $\lambda<0$.
The eigenvalue equation $(\lambda-a_\infty)u=0$ is a linear ordinary
differential equation of second order with four regular singular points $\rho=0,
\pm 1, \infty$.
However, applying the transformation $\rho \mapsto z:=\rho^2$ we can reduce this
number by $1$ and we are left with the three regular singular points
$z=0,1,\infty$.
Thus, solutions are given in terms of hypergeometric functions (cf.
\cite{Erdelyi1953}).
We define a new unknown 
$$v(z):=z^\alpha(1-z)^\beta u(\sqrt{z})$$ 
where
$\alpha:=-\frac{1}{4}(-1+i\sqrt{7})$ and 
$\beta:=-\frac{1}{2}(1+\sqrt{-\lambda})$.
The equation $(\lambda-a_\infty)u=0$ transforms into
$$
z(1-z) v''+ [c-(a+b+1)z]v'-ab v=0
$$
where $a:=\frac{1}{4}(1+2\sqrt{-\lambda}+i\sqrt{7})$, 
$b:=\frac{1}{4}(3+2\sqrt{-\lambda}+i\sqrt{7})$ and 
$c:=1+\frac{i\sqrt{7}}{2}$.
The solution $v_1(\cdot, \lambda)$ defined by $v_1(z, \lambda):=
{}_2F_1(a, b; a+b+1-c; 1-z)$ corresponds to the solution of $(\lambda-a_\infty)u=0$
which belongs to $H$ near $\rho=1$.
$v_1$ can be written as a linear combination of the form
$$ v_1(z, \lambda) \propto {}_2F_1(a,b; c; z)+m(\lambda)
z^{1-c}{}_2F_1(a+1-c,b+1-c; 2-c; z) $$
where the connection coefficient $m(\lambda)$ can be given in terms of the
$\Gamma$--function \cite{Erdelyi1953} and reads 
$$ m(\lambda)=\frac{\Gamma(a+1-c)\Gamma(b+1-c)\Gamma(c-1)}{\Gamma(a)
\Gamma(b)\Gamma(1-c)}. $$
Thus, the boundary condition at $\rho=0$ translates into $m(\lambda)=m(0)$.
This transcendental equation can be solved numerically and the results are given
in Table \ref{nlarge_tbl_ev}.

\begin{table}[h]
\begin{center}
\begin{tabular}{|c|c|c|c|c|c|}
\hline
 $\mu_n$ & $A_1$ & $A_2$ & $A_3$ & $A_4$ & $A_\infty$ \\
 \hline 
 $\mu_1$ & 5.333625 & 5.304 & 5.30 & 5.3 & 5.3009 \\ 
 $\mu_2$ & & 58.0701 & 57.68 & 57.6 & 57.637 \\
 $\mu_3$ & & & 625 & 620 & 619.61 \\
\hline
\end{tabular}
\end{center}
\caption{Approximate eigenvalues}
\label{nlarge_tbl_ev}
\end{table}
Thus, as conjectured, the point spectrum of $A_\infty$ seems to be the limit of
the point spectra of $A_n$.

\section{Generalization to higher equivariance indices}
Finally, we consider more general equivariant wave maps
described by the equation
\begin{equation}
\label{eq_hindex}
\psi_{tt}-\psi_{rr}-\frac{2}{r}\psi_r+\frac{\ell(\ell+1)}{2}
\frac{\sin(2\psi)}{r^2}=0 
\end{equation}
where $\ell \in \mathbb{N}$ is the equivariance index 
(see e.g. \cite{Shatah1994} for a derivation of this equation).
Self--similar solutions $\psi(t,r)=f(\frac{r}{T-t})$ of eq. (\ref{eq_hindex})
satisfy
$$ f''+\frac{2}{\rho}f'-\frac{\ell(\ell+1)}{2}\frac{\sin(2f)}{\rho^2
(1-\rho^2)}=0 $$ 
and again, we require $f(0)=0$ and $f(1)=\frac{\pi}{2}$ as regularity conditions.
Bizo\'n's proof of the existence of a countable set $\{f_{n,\ell}:
n=0,1,2\dots\}$ of
self--similar solutions carries over to this more general situation 
(see \cite{Bizon2000a}).
In fact, the solutions $f_{n,\ell}$ have analogous properties for all $\ell$. The index
$n$ counts the number of intersections of $f_{n,\ell}$ with the line $\frac{\pi}{2}$ in
the interval $\rho \in [0,1)$ and we have $f_{n,\ell}(\rho) \in (0,\pi)$ for all 
$n$ and
$\rho \in (0,1]$.
Furthermore, the asymptotic behaviour of $f_{n,\ell}$ for $\rho \to 0$ is 
$f_{n,\ell}(\rho)\sim \rho^\ell$.

Linear perturbations $\phi$ around $f_{n,\ell}$ satisfy the evolution equation
$$  \phi_{\sigma \sigma}-2 \phi_\sigma - (1-\rho^2)^2 \phi_{\rho
\rho}-\frac{2(1-\rho^2)^2}{\rho}\phi_\rho+\frac{\ell(\ell+1)(1-\rho^2)\cos(2
f_{n,\ell})}{\rho^2}\phi=0 $$
and analogous to the case $\ell=1$ we define the Sturm--Liouville expressions
$a_{n,\ell}$ by 
$$ a_{n,\ell}u:=\frac{1}{w}(-(pu')'+q_{n,\ell}u) $$
where $w(\rho):=\frac{\rho^2}{(1-\rho^2)^2}$, $p(\rho):=\rho^2$ and 
$$ q_{n,\ell}(\rho):=\frac{\ell(\ell+1)(1-\rho^2)\cos(2f_{n,\ell}(\rho))-\rho^2}{(1-\rho^2)^2}. $$
The Frobenius indices for the equation $a_{n,\ell}u=0$ are $\{-\ell-1, \ell\}$ at $\rho=0$ and
$\{\frac{1}{2},\frac{1}{2}\}$ at $\rho=1$ and thus, both endpoints are in the limit--point case.
It follows that the associated maximal operator $A_{n,\ell}$ on $H=L^2_w(0,1)$ 
is self--adjoint.
Again, the study of $A_{n,\ell}$ is equivalent to the spectral problem for
the radial Schr\"odinger operator and thus, the rest of the analysis
remains unchanged.
Hence, we have the following theorem.

\begin{theorem}
The spectrum of $A_{n,\ell}$ consists of a continuous part 
$\sigma_e(A_{n,\ell})=[0,\infty)$ and exactly $n$ negative eigenvalues. 
As a consequence,
$A_{0,\ell}$ is nonnegative, the Cauchy problem
$$ \left \{ \begin{array}{l}
\ddot{u}(\sigma)+A_{0,\ell} u(\sigma)=0 \mbox{ for } \sigma > 0\\
u(0)=u_0, \dot{u}(0)=u_1
\end{array} \right .
$$
for  $u_0,u_1 \in \mathcal{D}(A_{0,\ell})$ is well--posed and the real--valued function 
$\sigma \mapsto \|A_{0,\ell}^{1/2}
u(\sigma)\|_H^2+\|\dot{u}(\sigma)\|_H^2$ is constant for all $\sigma \geq 0$.
\end{theorem}

\section{Acknowledgments}

The authors would like to thank Piotr Bizo\'n and Horst R. Beyer for helpful
discussions.
This work has been supported by the Austrian Fonds zur F\"orderung der
wissenschaftlichen Forschung FWF Project P19126 and partly by the Fundacion
Federico.

\bibliography{paperdiss.bib}{}
\bibliographystyle{plain}

\end{document}